\documentclass[conference,10pt]{IEEEtran}
\usepackage[T1]{fontenc} %
\usepackage[utf8]{inputenc} %
\usepackage[english]{babel} %
\usepackage{amsmath}
\usepackage{amsthm}
\usepackage{amsfonts}
\usepackage{colonequals} %
\usepackage[cmintegrals]{newtxmath} %
\usepackage{mathtools} %
\usepackage{thmtools} %
\usepackage{textcomp}
\usepackage{stmaryrd} %
\usepackage{bm} %
\usepackage{siunitx} %
\usepackage{comment} %
\usepackage{microtype} %
\usepackage{fnpct} %
\usepackage{fontawesome5} %
\usepackage{csquotes} %
\usepackage[inline]{enumitem} %
\usepackage{tabularx} %
\usepackage{booktabs} %
\usepackage[hidelinks]{hyperref} %
\usepackage[capitalize,nameinlink]{cleveref} %
\usepackage{xcolor} %
\usepackage{graphicx} %
\usepackage[inkscapelatex=false]{svg} %
\usepackage{wrapfig} %
\usepackage{float}  %
\usepackage{placeins} %
\usepackage{xspace} %
\usepackage{balance}
\usepackage[noEnd]{algpseudocodex} %
\usepackage[
    backend=biber,
    style = ieee,
    citestyle = ieee-comp,
    sortlocale=en_US,
    sortcites=true,
    url=true,
    eprint=true,
    giveninits=false,
    dashed=false,
    minnames=1,
    maxnames=5
]{biblatex}
\AtEveryBibitem{%
    \clearname{editor}%
    \clearfield{series}%
    \clearfield{isbn}%
    \clearfield{issn}%
    \clearfield{day}%
    \clearfield{month}%
    \clearfield{place}%
    \clearlist{location}%
    \clearfield{pages}%
    \clearfield{volume}%
    \clearfield{number}%
    \ifentrytype{software}{%
    }{%
        \clearfield{url}%
        \clearfield{urlyear}%
        \clearfield{urldate}%
    }%
}
\DeclareSourcemap{
    \maps[datatype=bibtex]{
        \map{
            \step[fieldsource=doi, match=\regexp{.+/arXiv\..+}, final]
            \step[fieldsource=eprint, match=\regexp{.+}, final]
            \step[fieldset=doi, null]
        }
        \map{
            \step[fieldsource=doi, match=\regexp{.+}, final]
            \step[fieldset=eprint, null]
            \step[fieldset=archiveprefix, null]
            \step[fieldset=eprinttype, null]
        }
    }
}
\setlist[enumerate]{label=(\arabic*)} %
\makeatletter
\long\def\@makecaption#1#2{%
    \ifx\@captype\@IEEEtablestring%
    \footnotesize\bgroup\par\centering\@IEEEtabletopskipstrut{\normalfont\footnotesize {#1.}\nobreakspace\scshape #2}\par\addvspace{0.5\baselineskip}\egroup%
    \@IEEEtablecaptionsepspace
    \else
        \@IEEEfigurecaptionsepspace
        \setbox\@tempboxa\hbox{\normalfont\footnotesize {#1.}\nobreakspace #2}%
        \ifdim \wd\@tempboxa >\hsize%
        \setbox\@tempboxa\hbox{\normalfont\footnotesize {#1.}\nobreakspace}%
        \parbox[t]{\hsize}{\normalfont\footnotesize \noindent\unhbox\@tempboxa#2}%
        \else%
            \hbox to\hsize{\normalfont\footnotesize\hfil\box\@tempboxa\hfil}%
        \fi\fi}
\makeatother
\makeatletter
\let\MYcaption\@makecaption
\makeatother
\usepackage{subcaption} %
\makeatletter
\let\@makecaption\MYcaption
\makeatother
\graphicspath{ {figures/} }
\svgpath{ {figures/} }
\faStyle{regular}

\declaretheoremstyle[
    bodyfont=\itshape,%
]{example-style}
\declaretheorem[
    name=Example,%
    style=example-style,%
    numbered=unless unique,%
]{example}
\declaretheorem[
    name=Implementation,%
    style=example-style,%
    numbered=unless unique,%
]{implementation}
\crefname{section}{Sec.}{Sec.}
\Crefname{section}{Sec.}{Sec.}
\crefname{example}{Ex.}{Ex.}
\Crefname{example}{Ex.}{Ex.}
\newcolumntype{R}{>{\raggedleft\arraybackslash}X}
\newcolumntype{C}{>{\centering\arraybackslash}X}
\definecolor{TUM_blue}{RGB}{0,101,189}
\colorlet{TUM_black}{black}
\colorlet{TUM_white}{white}
\definecolor{TUM_darkblue}{RGB}{0,82,147}
\colorlet{TUM_darkblue100}{TUM_darkblue}
\colorlet{TUM_darkblue80}{TUM_darkblue100!80}
\colorlet{TUM_darkblue50}{TUM_darkblue100!50}
\colorlet{TUM_darkblue20}{TUM_darkblue100!20}
\definecolor{TUM_verydarkblue}{RGB}{0,51,89}
\colorlet{TUM_verydarkblue100}{TUM_verydarkblue}
\colorlet{TUM_verydarkblue80}{TUM_verydarkblue100!80}
\colorlet{TUM_verydarkblue50}{TUM_verydarkblue100!50}
\colorlet{TUM_verydarkblue20}{TUM_verydarkblue100!20}
\colorlet{TUM_darkgrey}{TUM_black!80}
\colorlet{TUM_grey}{TUM_black!50}
\colorlet{TUM_lightgrey}{TUM_black!20}
\definecolor{TUM_beige}{RGB}{218,215,203}
\definecolor{TUM_orange}{RGB}{227,114,34}
\definecolor{TUM_green}{RGB}{162,173,0}
\definecolor{TUM_verylightblue}{RGB}{152,198,234}
\definecolor{TUM_lightblue}{RGB}{100,160,200}
\newcommand{\ie}{i.\,e.\nolinebreak\@\xspace}

\newcommand{\eg}{e.\,g.\nolinebreak\@\xspace}

\renewcommand{\phi}{\varphi}

\newcommand{\set}[1]{\left\{#1\right\}}

\newcommand{\prop}[1]{{\normalfont\footnotesize\textsf{#1}}}
\usepackage{flushend} %
\definecolor{yannick}{HTML}{e89c1c}
\definecolor{robert}{HTML}{e5dd0c}
\definecolor{ludwig}{HTML}{76c11f}

\usepackage{pgfplots}
\pgfplotsset{compat=1.18}

\usepackage{listings}

\lstdefinestyle{python}{
    language=Python,
    morekeywords=[3]{%
        num_qubits,%
        duration,%
        error%
    }
}
\lstdefinelanguage{json}{
    keywords=[2]{true,false,null},
    string=[b]",
    comment=[l]{\#},
}
\lstdefinestyle{json}{
    language=json,
    stringstyle=\color{TUM_blue}
}
\title{Enabling Neutral Atom Integration: Redesigning\\Device Models for Universal Quantum Ecosystems}
\author{
    \IEEEauthorblockN{
        Yannick Stade\IEEEauthorrefmark{1},
        Lukas Burgholzer\IEEEauthorrefmark{1}\IEEEauthorrefmark{2},
        Ludwig Schmid\IEEEauthorrefmark{1},
        and Robert Wille\IEEEauthorrefmark{1}\IEEEauthorrefmark{2}
    }
    \IEEEauthorblockA{\IEEEauthorrefmark{1}%
    Chair for Design Automation,
        Technical University of Munich,
        Munich, Germany
    }
    \IEEEauthorblockA{\IEEEauthorrefmark{2}%
    Munich Quantum Software Company GmbH,
        Garching near Munich, Germany
    }
    \{%
    yannick.stade, %
    lukas.burgholzer, %
    ludwig.s.schmid, %
    robert.wille\}@tum.de\\
    \href{https://www.cda.cit.tum.de/research/quantum}{www.cda.cit.tum.de/research/quantum}
}
\hypersetup{ %
    pdftitle={Enabling Neutral Atom Integration: Redesigning Device Models for Universal Quantum Ecosystems},
    pdfsubject={IEEE International Conference on Quantum Computing and Engineering 2026},
    pdfauthor={
        Yannick Stade,
        Lukas Burgholzer,
        Ludwig Schmid,
        Robert Wille
    }
}
\begin{document}

    \maketitle

    \begin{abstract}
        Quantum computing is transitioning from an academic idea to a practical technology, driven by recent hardware advancements and clear paths toward real-world applications.
        Universal quantum ecosystems (e.g., Qiskit, Cirq, PennyLane) facilitate this transition by providing a consistent interface to diverse quantum devices, abstracting hardware-specific details through a \emph{device model} that captures each device's computational capabilities.
        However, these device models have historically been shaped by superconducting hardware, assuming static qubit positions and fixed coupling maps.
        This prevents them from representing the unique computational capabilities of emerging technologies such as neutral atoms, which feature dynamic qubit rearrangement and zoned operations.
        As a result, although numerous specialized compilers for neutral atom devices already exist, they cannot retrieve the hardware information they need through these ecosystems---creating a technology lock that hinders or even prevents the integration of neutral atom devices.
        In this work, we demonstrate how this limitation leads to suboptimal compilation results and can exclude certain devices entirely.
        Motivated by this, we propose rethinking current device models to faithfully represent neutral atom devices, enabling their seamless integration into universal quantum ecosystems.
        Evaluations conducted within the Quantum Device Management Interface (QDMI) demonstrate that the proposed device model unlocks a routing overhead fidelity improvement by a factor of up to 100,000 on a circuit with 16 qubits and 600 gates.
    \end{abstract}


    \section{Introduction}\label{sec:introduction}

    \emph{Universal quantum ecosystems} provide a uniform interface to access different quantum devices, allowing users to develop quantum applications without worrying about the underlying hardware details.
    This abstraction is crucial for the broader adoption of quantum computing, as it allows developers to focus on algorithm design rather than hardware-specific implementation details.
    To this end, multiple such ecosystems have emerged over the past few years, with arguably the most prominent example being Qiskit~\cite{qiskit2024}, which was developed to support the quantum devices provided by IBM~\cite{ibmQuantumExperience}.
    Other ecosystems such as Cirq~\cite{Cirq_Developers_Cirq_2025}, PennyLane~\cite{bergholm2022pennylaneautomaticdifferentiationhybrid}, and initiatives such as the Munich Quantum Valley~\cite{munichQuantumValleyKicks2021}, the Quantum Delta Delft~\cite{quantumInsiderLaunchingQuantumDelta2025}, the Chicago Quantum Exchange~\cite{uchicagoArgonneFermilabPrepare2017}, Riken Quantum~\cite{rikenEstablishesCenterQuantum2021}, etc., have gained significant traction, supporting devices from IBM, Google, AWS, Rigetti, IQM, IonQ, and others.

    A key component of these ecosystems is their \emph{device model}, abstracting the quantum device's \emph{computational capabilities} such as qubit count, native gate set, and operation durations and fidelities.
    Compilers consume this information to transform high-level (hardware-agnostic) quantum circuits into low-level (hardware-specific) instructions.

    However, existing device models were primarily designed for \emph{superconducting}~(SC) qubits---the most widely available technology to date~\cite{aruteQuantumSupremacyUsing2019,ibmQuantumExperience}.
    While they can abstract arbitrary topologies, they assume fixed qubit positions and describe connectivity through a static coupling map.
    Meanwhile, other technologies such as \emph{neutral atoms}~(NAs,~\cite{bluvsteinLogicalQuantumProcessor2023,bluvsteinArchitecturalMechanismsUniversal2025}) have emerged, offering unique advantages like scalability, long coherence times, and dynamic qubit rearrangement---features that fundamentally break the static coupling assumption and cannot be abstracted within current device models.

    Over the past few years, numerous compilation methods specifically tailored to NAs have been proposed~\mbox{\cite{patelGeyserCompilationFramework2022,patelGRAPHINEEnhancedNeutral2023,ludmirPARALLAXCompilerNeutral2024,wangQPilotFieldProgrammable2024,silverQomposeTechniqueSelect2024,tanCompilationDynamicallyFieldProgrammable2025,tanCompilingQuantumCircuits2024,wangAtomiqueQuantumCompiler2024,nottinghamDecomposingRoutingQuantum2023,stadeAbstractModelEfficient2024,schmidHybridCircuitMapping2024,stadeRoutingAwarePlacementZoned2025}}, exploiting features like dynamic qubit rearrangement and native multi-qubit gates.
    However, none of these compilers have been integrated into universal quantum ecosystems, because the underlying device models cannot abstract the NA-specific information these compilers require.
    This creates a \emph{gap}: in scalable quantum computing environments---such as those at HPC centers~\cite{munichQuantumValleyKicks2021}---where devices and compilers should be exchangeable in a plug-and-play fashion, neither NA devices nor their specialized compilers can be integrated.
    The missing piece is not yet another NA compiler, but rather a device model that can bridge the gap between NA hardware and the software ecosystem with a standardized interface.

    In this work, we address this gap: we analyze the limitations current device models impose, demonstrating in \cref{sec:motivation} how they hinder the utilization of NA devices.%
    Motivated by this, we rethink existing device models in \cref{sec:approach}, proposing targeted extensions that capture the unique computational capabilities of NAs.%
    In \cref{sec:demonstration}, we implement the resulting more generic device model on top of the Quantum Device Management Interface~(QDMI,~\cite{willeQDMIQuantumDevice2024}).%
    Using one representative NA compiler, we demonstrate that by bridging the abstraction gap with the more generic device model, the routing overhead fidelity improves by a factor of up to 100,000 on a circuit with 16 qubits and 600 gates.

    \section{Preliminaries}\label{sec:background}
    To understand why existing device models struggle to represent NA devices, this section reviews:
    \begin{enumerate*}
        \item fundamentals of NA-based quantum computing and
        \item existing device models in universal quantum ecosystems.
    \end{enumerate*}

    \subsection{Neutral Atom-based Quantum Computing}\label{subsec:neutral atom}

    \emph{Neutral atoms}~(NAs,~\cite{bluvsteinLogicalQuantumProcessor2023,bluvsteinArchitecturalMechanismsUniversal2025}) have emerged as a promising quantum computing platform.
    They differ significantly from more established \emph{superconducting}~(SC) qubits~\cite{schmidComputationalCapabilitiesCompiler2024}.
    Following, we review the setup and the native operations of NAs. %

\begin{figure}[t]
        \centering
        \includegraphics[width=\linewidth]{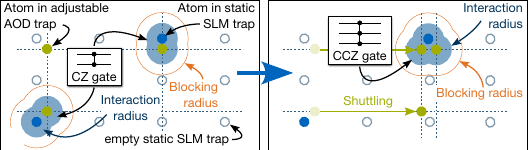}\\
        \raggedright\vspace{-16pt}\hspace{2pt}%
        \begin{subfigure}{128pt}
            \caption{\raggedright}
            \label{subfig:background a}
        \end{subfigure}%
        \begin{subfigure}{30pt}
            \caption{\raggedright}
            \label{subfig:background b}
        \end{subfigure}\\
        \vspace{-6pt}
        \caption{\textbf{(a)}~two atoms (blue) in SLM traps; three atoms (green) in AOD traps; two atom pairs perform a CZ gate. \textbf{(b)}~two leftmost AOD atoms were shuttled right; a group of three atoms performs a CCZ gate.}
        \label{fig:background}
        \vspace{-15pt}
    \end{figure}

    First, qubits are encoded in the electronic states of neutral atoms from the group of alkali or alkaline-earth metals.
    The atoms are confined in space by optical tweezers or lattices and cooled down to their motional ground state~\cite{barredoAtombyatomAssemblerDefectfree2016}.
    These trap sites define possible locations of atoms; in particular, not all sites need to be occupied. %
    A static trapping of atoms can be realized using \emph{Spatial Light Modulators}~(SLMs,~\cite{bluvsteinQuantumProcessorBased2022}).
    \begin{example}
        \Cref{fig:background} shows trap sites (grey circles) arranged as a grid, which can be occupied by atoms (blue).
    \end{example}

    Second, one-qubit operations are realized through \mbox{laser-driven} electronic state transitions, while \mbox{multi-qubit} operations use the principle of the Rydberg blockade to implement controlled phase gates~\cite{bluvsteinQuantumProcessorBased2022,everedHighfidelityParallelEntangling2023,grahamRydbergMediatedEntanglementTwoDimensional2019}.
    This can be described by an \emph{interaction radius}~$r_\mathrm{int}$ and a \emph{blocking radius}~$r_\mathrm{b} > r_\mathrm{int}$~\cite{schmidComputationalCapabilitiesCompiler2024}.
    When illuminated by a Rydberg laser, all atoms within $r_\mathrm{int}$ perform a multi-qubit phase gate.
    However, simultaneously, all illuminated atoms located within $r_\mathrm{b}$ perturb each other. %
    Thus, the \emph{qubit connectivity} depends on atom positions and changes with rearrangement, unlike the fixed coupling map in SC devices.
    Furthermore, by accumulating multiple atoms, this also allows the direct native execution of multi-controlled phase gates such as the CCZ gate.
    \begin{example}
        \Cref{fig:background} illustrates CZ and CCZ gates by arranging atoms within $r_\mathrm{int}$ (blue area).
        Note that no other atoms are located within the $r_\mathrm{b}$ (orange).
    \end{example}

    Third, many NA devices allow the \emph{dynamic rearrangement}, so-called \emph{shuttling}, of atoms during computation using \emph{Acousto-Optic Deflectors}~(AODs,~\cite{bluvsteinQuantumProcessorBased2022}).
    When satisfying certain constraints~\cite{stadeAbstractModelEfficient2024}, this facilitates arbitrary rearrangements of atoms, constituting a promising routing alternative to SWAP gate insertion~\cite{schmidHybridCircuitMapping2024}.
    This requires fundamentally new compilation methods~\mbox{\cite{stadeRoutingAwarePlacementZoned2025,stadeAbstractModelEfficient2024,tanCompilingQuantumCircuits2024,wangAtomiqueQuantumCompiler2024}} targeted to NA-specific computational capabilities, which are incompatible with common universal quantum ecosystems.
    \begin{example}
        The transition from \cref{subfig:background a} to \cref{subfig:background b} illustrates shuttling, moving the two green, leftmost atoms to the right.
    \end{example}

    Finally, most operations on NAs can be applied either locally by focusing the corresponding laser beam onto a specific atom site or globally over a certain region.
    We call the latter \emph{zoned operations}.
    This allows the design of \emph{zoned devices}~\mbox{\cite{bluvsteinQuantumProcessorBased2022,stadeAbstractModelEfficient2024}} where spatially separated \emph{zones} are optimized for certain operations.
    Typically, these include a \emph{storage zone} with long coherence times, an \emph{entanglement zone} which is illuminated by a Rydberg beam, and a separate \emph{measurement zone} to read out qubit states without disturbing other atoms.
    While this concept of zoned operations also exists for trapped-ion hardware~\cite{muraliArchitectingNoisyIntermediateScale2020}, it is not native to SC-chips and, therefore, not incorporated in common device models.
    
    \subsection{Abridged History of Device Models}\label{subsec:device model}

    The launch of IBM's five-qubit SC device in 2016~\cite{ibmQuantumExperience} marked a pivotal moment in the field.
    It necessitated the development of quantum ecosystems that could bridge the gap between \mbox{high-level} quantum algorithms and quantum devices.
    Qiskit~\cite{qiskit2024}, released alongside IBM Quantum Platform, quickly became a widely adopted open-source quantum ecosystem, evolving to support multiple quantum devices.
    To provide a uniform user experience, \ie, to be able to switch between different devices without changing the code, Qiskit introduced a compilation pipeline that transforms hardware-agnostic circuits into device-specific instructions.
    To adapt to diverse devices, this requires a standardized representation of each device's computational capabilities---the \emph{device model}.
    Specifically, Qiskit's device model captures the number of qubits, the native gate set with corresponding properties like duration and fidelity for each qubit, and a coupling map defining the qubit's connectivity.

    \begin{example}
        The following Python code shows a \texttt{Target} (\ie, Qiskit's device model) for a two-qubit device that supporting a \texttt{UGate} on both qubits and a \texttt{CXGate} between them.
    \end{example}
    \begin{lstlisting}[style=python]
from qiskit.transpiler import Target, InstructionProperties
from qiskit.circuit.library import UGate, CXGate
# Create a new target
targ = Target(num_qubits=2)
# Define properties for UGate on both qubits
u_props = {
    (0,): InstructionProperties(duration=5e-8, error=0.0001),
    (1,): ... # same as above
}
targ.add_instruction(UGate(0, 0, 0), u_props)
# Properties for CXGate between qubits 0 and 1
cx_props = {
    (0, 1): InstructionProperties(duration=5e-7, error=0.005),
}
target.add_instruction(CXGate(), cx_props)
    \end{lstlisting}

    Following Qiskit's lead, several other quantum ecosystems introduced their own device models.
    PennyLane~\cite{bergholm2022pennylaneautomaticdifferentiationhybrid} employs a \texttt{qml.device} class. %
    While Qiskit's \texttt{Target} is merely a description of the device, \texttt{qml.device} allows users to execute circuits and query device properties, such as the native gate set and qubit connectivity, in a backend-agnostic manner.
    Cirq~\cite{Cirq_Developers_Cirq_2025}, Google's quantum computing framework, introduces a \texttt{cirq.Device} class, providing functions to validate circuits against the device's constraints. %
    More recently, \mbox{CUDA-Q}~\cite{cuda-q} has defined its own device model that allows for the representation of different \emph{quantum processing units}~(QPUs).
    Despite differences in implementation, all these models share a common heritage: while they can represent arbitrary static qubit topologies, they were fundamentally shaped by SC qubits and do not natively capture dynamic rearrangement, spatial qubit coordinates, or zoned operations---capabilities essential for emerging platforms like NA devices.

    \section{Limitations of Existing Device Models}\label{sec:motivation}

    Universal quantum ecosystems rely on device models to provide a consistent user experience across diverse quantum devices.
    Only then can users seamlessly switch between devices without changing their application code by simply providing another device model instance.
    However, the limitations of existing device models result in significant drawbacks when attempting to utilize existing universal quantum ecosystems for emerging technologies, such as NAs.
    There are multiple computational capabilities of NAs missing in existing device models, \eg, the dynamic rearrangement of atoms and zoned operations.
    This leads to suboptimal compilation results, prevents the usage of certain NA devices entirely, and blocks the integration of specialized NA compilers into these ecosystems.

    \subsection{Untapped Potential of Dynamic Rearrangement}\label{subsec:untapped monolithic}

    Existing device models, designed for SC devices, only capture a fixed coupling map and a set of native gates together with their fidelities and durations.
    NA devices, in contrast, enable dynamic qubit rearrangement via shuttling, as reviewed in \cref{subsec:neutral atom}.
    To leverage this powerful computational capability, compilers must know the current and target locations of atoms to generate a valid schedule of shuttling operations that respects the physical constraints mentioned in \cref{subsec:neutral atom}---information that the compiler cannot retrieve from existing device models.

    \begin{example}\label{exp:motivation monolithic}
    Consider the circuit in \cref{subfig:motivation circuit} that realizes a state preparation of a logical qubit in the \(|0\rangle\)-state using the Steane code~(cf. \cref{subig:motivation steane},~\cite{steaneFastFaulttolerantFiltering2004,stadeOptimalStatePreparation2024}).
    This circuit results in the required interactions depicted in \cref{subfig:motivation graph}.
    For NA devices supporting local CZ gates, connectivity---as long as qubits are not rearranged---can be represented by a static coupling map, a computational capability covered by \mbox{state-of-the-art} device models.
    \Cref{subfig:motivation swap} depicts the coupling map of an NA device allowing for nearest-neighbor interactions as a light blue grid.

    \begin{figure}[t]
        \centering
        \includegraphics[width=0.9\linewidth]{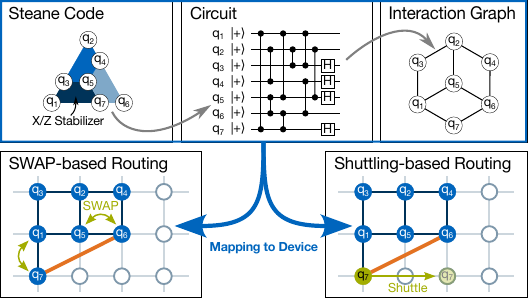}\\
        \raggedright\vspace{-82pt}\hspace{14pt}%
        \begin{subfigure}{78pt}
            \caption{\raggedright}
            \label{subig:motivation steane}
        \end{subfigure}%
        \begin{subfigure}{86pt}
            \caption{\raggedright}
            \label{subfig:motivation circuit}
        \end{subfigure}%
        \begin{subfigure}{30pt}
            \caption{\raggedright}
            \label{subfig:motivation graph}
        \end{subfigure}\\
        \vspace{52pt}\hspace{14pt}%
        \begin{subfigure}{140pt}
            \caption{\raggedright}
            \label{subfig:motivation swap}
        \end{subfigure}%
        \begin{subfigure}{80pt}
            \caption{\raggedright}
            \label{subfig:motivation shuttling}
        \end{subfigure}\\
        \vspace{-7pt}
        \caption{\textbf{(a)}~the six stabilizers of the Steane code. \textbf{(b)}~a circuit preparing its logical \(|0\rangle\)-state with NA-native gates. \textbf{(c)}~the resulting interaction graph (edges are CZ gates). \textbf{(d)}~a qubit mapping to an NA device featuring local addressable multi-qubit gates. At least one CZ gate (the orange connection) requires the insertion of SWAP gates to route qubits. \textbf{(e)}~the missing connection is enabled by shuttling offered by NA devices.}
        \label{fig:motivation-monolithic}
        \vspace{-15pt}
    \end{figure}

    Since the circuit's interaction graph, cf.~\cref{subfig:motivation graph}, is not isomorphic to any subgraph of the coupling map, missing connections have to be compensated for via SWAP gate insertions~\cite{zulehnerEfficientMethodologyMapping2019}, as illustrated in \cref{subfig:motivation swap}.
    This approach, however, significantly degrades the overall fidelity: one SWAP gate, which is decomposed into three CZ gates (besides \mbox{single-qubit} gates), leads to a routing overhead with an additional infidelity of at least 0.15 (given a fidelity of 99.5\% per local CZ gate~\cite{bluvsteinLogicalQuantumProcessor2023}).

    This SWAP-based routing leaves substantial potential of the NA device untapped.
    In fact, the missing connection can alternatively be established by rearranging one atom via shuttling, as illustrated in \cref{subfig:motivation shuttling}.
    In contrast to SWAP gate execution, shuttling can be performed with a significantly higher fidelity: Assuming a mean shuttling speed of \SI{0.55}{\micro\meter\per\micro\second} and an atom separation of \SI{3}{\micro\meter} results in a time overhead of \SI{10.9}{\micro\second} to move atom \(q_7\) two sites right~\cite{bluvsteinLogicalQuantumProcessor2023}.
    This routing overhead corresponds to an infidelity of only \(6.5 \times 10^{-5}\) when using the following formula to approximate the coherence and defining \(t_\mathrm{idle} = n_\mathrm{atoms} \cdot \SI{10.9}{\micro\second}\) and \mbox{\(T_\mathrm{eff} = \SI{1.5}{\s}\)~\cite{bluvsteinLogicalQuantumProcessor2023,schmidComputationalCapabilitiesCompiler2024}:}\vspace{-4pt}
    \begin{align}
        \exp\left(-t_\mathrm{idle} / T_\mathrm{eff}\right)\enspace.\label{eq:coherence}
    \end{align}\vspace{-16pt}\\\noindent
    Hence, because universal quantum ecosystems rely on device models that do not properly represent the capabilities of NA devices, they yield the circuit execution in \cref{subfig:motivation swap} rather than the one in \cref{subfig:motivation shuttling}, leading to a significant performance degradation.
    \end{example}

    The example clearly demonstrates that using the additional computational capabilities of NAs can significantly improve the performance of the executed circuits.
    However, due to the lack of adequate device models, this potential remains untapped in existing universal quantum ecosystems.

    \subsection{Exclusion of Zoned Devices}\label{subsec:inaccessible zoned}
    In addition to the performance degradation above, the limitations of current device models can even exclude certain NA devices entirely from universal quantum ecosystems.

    \begin{figure}[b]
        \vspace{-12pt}
        \centering
        \includegraphics[width=0.97\linewidth]{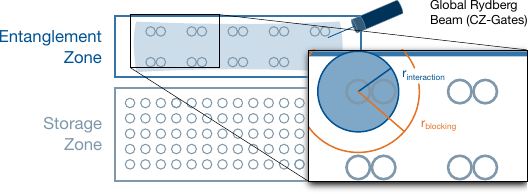}\\
        \raggedright\vspace{-22pt}\hspace{2pt}%
        \begin{subfigure}{148pt}
            \caption{\raggedright}
            \label{subfig:motivation zones model}
        \end{subfigure}%
        \begin{subfigure}{30pt}
            \caption{\raggedright}
            \label{subfig:motivation zones interaction}
        \end{subfigure}\\
        \vspace{-6pt}
        \caption{\textbf{(a)}~schematic of a zoned NA device showing the storage and entanglement zone; \textbf{(b)}~magnification of the entanglement zone depicting the interaction and blocking radius clearly demonstrating that swapping qubits between different site pairs is infeasible.}
        \label{fig:motivation zones}
    \end{figure}

    \begin{example}\label{exp:motivation zones}
        Consider a zoned NA device as illustrated in \cref{subfig:motivation zones model}, where the storage zone is optimized for long coherence times and CZ gates can only be executed as zoned operations in the entanglement zone~\cite{bluvsteinLogicalQuantumProcessor2023}.
        This setup---featuring better coherence times~\mbox{\cite{grahamMultiqubitEntanglementAlgorithms2022,everedHighfidelityParallelEntangling2023}} compared to the monolithic device in \cref{exp:motivation monolithic}---depends on shuttling atoms between zones.
        The interaction radius depicted in \cref{subfig:motivation zones interaction} demonstrates that SWAP-based routing is infeasible, as atoms can only interact with their direct neighbors. %
        Consequently, compilation methods that rely on a device model only capturing a static coupling map, and, hence, are limited to SWAP-based routing, cannot target such zoned NA devices at all.
    \end{example}

    Both examples demonstrate that the missing computational capabilities of NA devices in existing device models create a substantial barrier to progress.
    Importantly, the solution to this barrier is not yet another NA-specific compiler---those already exist~\mbox{(cf.\ \cref{sec:introduction})}---but rather a device model that can convey all necessary hardware information to these compilers within a universal quantum ecosystem.

    \section{Device Model for Neutral Atom Devices}\label{sec:approach}

    We propose targeted extensions to existing device models that address the limitations discussed above, thereby enabling NA devices and the specialized compilers developed for them to be seamlessly integrated into universal quantum ecosystems.
    As one representative realization, we build upon the \emph{Quantum Device Management Interface}~(QDMI,~\cite{willeQDMIQuantumDevice2024}) employed in large universal quantum ecosystems.
    Note that, despite the concrete implementation, the underlying concepts are general and can be applied to any other device model, including those reviewed in \cref{subsec:device model}.
    This section briefly reviews QDMI and introduces the corresponding contributions.

    \subsection{Running Example: QDMI}\label{subsec:qdmi}

    QDMI abstracts over various hardware technologies and provides a uniform interface for devices. %
    It defines three different parties that closely interact with each other:
    \begin{enumerate*}
        \item the \emph{device} representing the actual quantum hardware,
        \item the \emph{client}, a software component utilizing the device, \eg, a mapping pass, and
        \item the \emph{driver} managing the communication between multiple devices and clients.
    \end{enumerate*}
    To retrieve information from a device, the client sends a corresponding query to the driver, which in turn forwards the query to the respective device and returns the received answer to the client.
    Properties can be queried for three different entities defined in QDMI's device model:
    \begin{enumerate*}
        \item for the \emph{device} itself, \eg, its qubit count,
        \item for \emph{sites} denoting possible qubit locations (equals qubits in case of SC), and
        \item for \emph{operations}, \eg, their current fidelity.
    \end{enumerate*}
    Each of those three entities has its own list of pre-defined properties.
    \begin{example}\label{exp:qdmi coupling map}
        The \prop{coupling map} is a property of the device.
        Querying the device for its coupling map returns a list of site pairs, where each pair indicates that a two-qubit operation can be executed between the two sites.
    \end{example}
    Similar to other universal quantum ecosystems, QDMI's device model is significantly influenced by SC devices and currently does not effectively capture the specific computational capabilities of NA devices.

    \subsection{Multi-Qubit Operations}\label{subsec:multi-qubit}

    In contrast to SC, where qubits have a fixed position and interaction partners are given by a fixed coupling map, NAs feature dynamic rearrangement.
    Similarly, an NA device defines potential atom locations through trap sites rather than a fixed qubit layout.
    The ability to undergo a multi-qubit gate is then determined via an interaction and blocking radius, as reviewed in~\cref{subsec:neutral atom}---both of which are missing in existing device models.
    Moreover, to determine whether two atoms are within the interaction radius, their precise location must be known.
    Hence, the device model must communicate the location of its sites, which can be in two- or three-dimensional space.

    \begin{implementation}\label{impl:multi-qubit}
    To facilitate multi-qubit operations while allowing dynamic rearrangement, the device model must be equipped with the following additional site properties:
    \begin{itemize}
        \item \prop{X-, Y-, Z-coordinate}: precise location of each site in space, whereof the Z-coordinate is optional
        \item \prop{interaction radius}: distance within which atoms can interact
        \item \prop{blocking radius}: distance within which atoms perturb each other during multi-qubit operations
    \end{itemize}
    \end{implementation}

    \subsection{Topology of Sites}\label{subsec:topology}

    The arrangement of trap sites on NA devices commonly follows a grid structure, but arbitrary geometries are possible, while only regular geometries are meaningful in the context of large-scale NA quantum computing.
    For a client to retrieve all sites, it can query the device for the already existing \prop{sites} property. %
    To construct the list of sites in the device implementation, we use the following systematic method.

\begin{figure}[b]
        \vspace{-6pt}
        \centering
        \includegraphics[width=\linewidth]{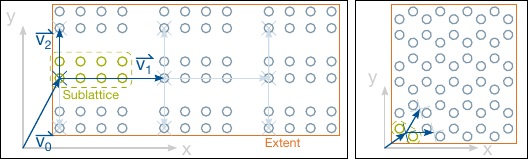}\\
        \raggedright\vspace{-16pt}\hspace{2pt}%
        \begin{subfigure}{170pt}
            \caption{\raggedright}
            \label{subfig:lattice grid}
        \end{subfigure}%
        \begin{subfigure}{30pt}
            \caption{\raggedright}
            \label{subfig:lattice honeycomb}
        \end{subfigure}\\
        \vspace{-6pt}
        \caption{(a)~a group of sites, the sublattice in green, is duplicated and shifted by two lattice vectors \(\overset{\rightharpoonup}{v_1}\) and \(\overset{\rightharpoonup}{v_2}\) in blue to form a 2D lattice of sites within a given extent in orange; (b)~also other structures like honeycomb lattices can be defined using this method.}
        \label{fig:lattice}
    \end{figure}

    Assuming the device provides sites in a two-dimensional lattice structure, as illustrated in~\cref{subfig:lattice grid}, where a group of eight sites, the \emph{sublattice} (green), is repeated horizontally and vertically.
    The vector \(\overset{\rightharpoonup}{v_0}\) denotes the sublattice's origin relative to the device's origin.
    The sublattice itself is defined as a list of eight coordinates relative to the sublattice's origin.
    It is then repeated by shifting it by multiples of the vectors \(\overset{\rightharpoonup}{v_1}\) and \(\overset{\rightharpoonup}{v_2}\).
    A rectangular extent (orange) limits this definition of sites to a finite number.
    \Cref{subfig:lattice honeycomb} shows how a honeycomb structure can be defined using the same method.
    Here, the sublattice consists of two sites, and the basis vectors are chosen accordingly.

    Based on the outlined method, different zones can easily be defined as different lattices, \eg, a storage zone with a large extent and dense site arrangement and a smaller entanglement zone with a sparser arrangement of site pairs.
    The sites themselves are then still passed from the device to the client as a plain list.
    However, in some clients, \eg, a mapping pass, it will be beneficial to recover this construction of sites for a more efficient representation of the device.

    \begin{implementation}
        To keep the necessary modifications to the existing device model minimal, we introduce the following additional site properties that guide the client in reconstructing the lattice structure: %
        \begin{itemize}
            \item \prop{module index} identifies a lattice (\ie, zone)
            \item \prop{submodule index} identifies a sublattice within a lattice
        \end{itemize}
    \end{implementation}

    \subsection{Dynamic Rearrangement}\label{subsec:shuttling}
    One key computational capability of NA devices is the dynamic rearrangement of atoms.
    The device can report shuttling, \emph{loading} (SLM \(\rightarrow\) AOD), and \emph{storing} (AOD \(\rightarrow\) SLM) as supported operations, alongside single- and multi-qubit gates.
    Most of their properties are already covered by existing operation properties, \eg, fidelity and duration.

    \begin{implementation}
        Nevertheless, some additional operation properties are necessary to fully capture the constraints of rearrangement operations.
        \begin{itemize}
            \item \prop{mean shuttling speed}: average atom movement speed
            \item \prop{minimal atom distance}: minimum atom separation during shuttling
        \end{itemize}
    \end{implementation}

    \subsection{Zoned Operations}\label{subsec:zones}

    An operation can report the sites it is applicable to by querying the property \prop{sites} on an operation, which fully suffices for local operations on NA devices.
    However, zoned operations act on all atoms in a region, including those held in adjustable AOD traps, not just on individual sites.
    For this purpose, we introduce a new type of sites, the \emph{zone sites}. %
    While regular sites define a point in space without an extent, zone sites have a two- or three-dimensional extent.

    \begin{implementation}
        Hence, we introduce the corresponding site properties:
        \begin{itemize}
            \item \prop{is zone}: a boolean property indicating whether this site is a regular or a zone site
            \item \prop{X-, Y-, Z-extent}: the zone's extent (the site's origin is denoted by the coordinates introduced above)
        \end{itemize}
        Additionally, we extend the operation properties:
        \begin{itemize}
            \item \prop{is zoned}: a boolean property indicating whether this operation is a zoned one
            \item \prop{idling fidelity}: only relevant for zoned multi-qubit gates, it denotes their fidelity on isolated atoms without an interaction partner that undergo a (faulty) identity operation~\cite{stadeOptimalStatePreparation2024}
        \end{itemize}
    \end{implementation}

    \subsection{Length and Duration Units}\label{subsec:units}
    The new properties comprise a couple of length values, whose unit must be specified for correct interpretation.
    For the device model to be able to choose its own unit length (and unit duration), the following two properties are added to the device.
    This approach has the central advantage that fixed-precision integers can be used for length and duration values, avoiding any precision loss due to rounding errors when calculating with length values.

    \begin{implementation}
        Hence, we equip the device with one further property to report the unit length of the device and apply the same method for durations:
        \begin{itemize}
            \item \prop{length scale factor}: integer denoting the unit length as multiples of \SI{1}{\nano\meter}
            \item \prop{duration scale factor}: scale factor for duration values as multiples of \SI{1}{\nano\second}
        \end{itemize}
        As a consequence, all duration and length values are returned as signed or unsigned integers.
    \end{implementation}

    \section{Demonstrating Unlocked Potential}\label{sec:demonstration}

    We implemented the approach proposed above.
    The corresponding full code is publicly available as part of the \emph{Munich Quantum Toolkit}~(MQT,~\cite{willeMQTHandbookSummary2024}) at \url{https://github.com/munich-quantum-toolkit/core}.
    By that, we provide a device model that fully captures the computational capabilities of NA devices.
    This enables specialized NA compilers to retrieve all necessary hardware information through the standardized interface, facilitating their seamless integration into universal quantum ecosystems.
    In the following, \cref{subsec:device implementation} first briefly outlines the implementation.
    Afterward, \cref{subsec:device spec} demonstrates how to use the device model.
    Finally, in \cref{subsec:case study}, we demonstrate how using the resulting device model fully unlocks the computational capabilities of NA devices and leads to improvements in the routing overhead fidelity by a factor of up to 100,000 on a circuit with 16 qubits and 600 gates.

    \subsection{Implementation of the Proposed Device Model}\label{subsec:device implementation}

    \begin{figure}[t]
        \centering
        \includegraphics[width=\linewidth]{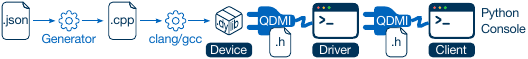}\\
        \raggedright\vspace{-4pt}\hspace{3pt}%
        \begin{subfigure}{51pt}
            \caption{\raggedright}
            \label{fig:qdmi flow:a}
        \end{subfigure}%
        \begin{subfigure}{52pt}
            \caption{\raggedright}
            \label{fig:qdmi flow:b}
        \end{subfigure}%
        \begin{subfigure}{50pt}
            \caption{\raggedright}
            \label{fig:qdmi flow:c}
        \end{subfigure}%
        \begin{subfigure}{56pt}
            \caption{\raggedright}
            \label{fig:qdmi flow:d}
        \end{subfigure}%
        \begin{subfigure}{30pt}
            \caption{\raggedright}
            \label{fig:qdmi flow:e}
        \end{subfigure}\\
        \vspace{-8pt}
        \caption{A JSON specification of the NA device is compiled into a shared library with QDMI as an interface. The library is loaded and accessed by the driver. Our driver implementation provides Python bindings that interact with the driver and eventually with the device.}
        \label{fig:qdmi flow}
        \vspace{-12pt}
    \end{figure}

    A QDMI~\cite{willeQDMIQuantumDevice2024} device implementation is a shared library with a well-defined C interface.
    To ease the definition of such a library, we decided to first specify the device's computational capabilities in a JSON file, depicted in \cref{fig:qdmi flow:a}.
    Then, this JSON specification is parsed by a custom generator that produces C++ source code as shown in \cref{fig:qdmi flow:b}.
    Finally, this source code is compiled into the shared library, shown in \cref{fig:qdmi flow:c}.
    The shared device library is then loaded by a QDMI driver, illustrated in \cref{fig:qdmi flow:d}.
    It communicates with the shared library through QDMI, illustrated as the blue plug in \cref{fig:qdmi flow:d}.
    Complementary, it provides the interface to clients, depicted in \cref{fig:qdmi flow:e}.
    For ease of use, our driver also provides Python bindings, allowing clients to access the device through Python.

    \subsection{JSON Specification and Usage}\label{subsec:device spec}

    \begin{figure}[b]
        \centering
        \includegraphics[width=\linewidth]{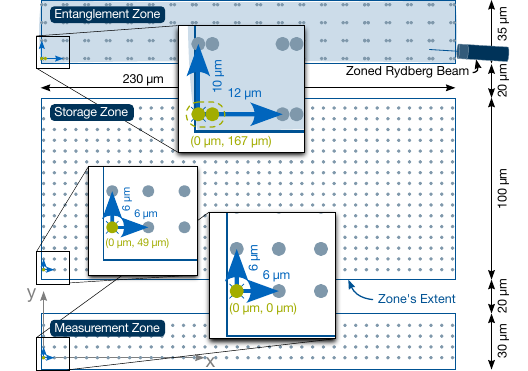}\\
        \vspace{-8pt}
        \caption{Schematic of the zoned NA device presented in~\cite{bluvsteinLogicalQuantumProcessor2023}.}
        \label{fig:architecture}
    \end{figure}

    For our demonstration, we model the NA device presented in~\cite{bluvsteinLogicalQuantumProcessor2023} and illustrated in~\cref{fig:architecture}, which features an entanglement, storage, and measurement zone.
    The figure also depicts the dimensions of the zones and the distances between the sites.
    The following code snippet shows an excerpt of the corresponding JSON specification.
    The visible part defines the entanglement zone according to the lattice structure explained in \cref{subsec:topology}.

    \begin{lstlisting}[style=json]
"lengthUnit": {"unit": "um", "scaleFactor": 1.},
"traps": [{
  "latticeOrigin":  {"x": 0, "y": 167},
  "latticeVector1": {"x": 12, "y": 0},
  "latticeVector2": {"x": 0, "y": 10},
  "sublatticeOffsets": [{"x": 0, "y": 0},
                        {"x": 2, "y": 0}],
  "extent": {"origin": {"x": 0, "y": 164},
             "size":   {"width": 230, "height": 35}
  }, ... # other zones
}], ... # other device properties
    \end{lstlisting}

    Afterward, we can reconstruct the lattice structure in a client by querying the device solely for properties defined in the device model, including the newly introduced ones.
    In the following pseudocode, \(v_{0,1,2}\) stand for the lattice origin, vector~1, and vector~2, respectively.
    Furthermore, \(O\) denotes the set of offsets defining the sublattice, and \(R\) is an auxiliary set of reference sites for each sublattice.
\vspace{-3pt}

\noindent\rule{\linewidth}{0.4pt}\vspace{-4pt}
\begin{algorithmic}
    \ForAll{\prop{module indices}}
        \State \(v_0 \gets \min\set{(x, y)\mid \text{sites with min. \prop{submodule index}}}\)
        \State \(O \gets \{(x - x(v_0), y - y(v_0))\mid\) \\\hspace*{2cm}sites with min. \prop{submodule index}\(\}\)
        \State \(R \gets \{\min\set{(x, y)\mid \text{sites with \prop{submodule index} } i}\) \\\hspace*{2cm}for all \prop{submodule indices} \(i\}\)
        \State \(v_1 \gets (x - x(v_0), y - y(v_0))\) for \((x, y)\in R\) closest to \(v_0\)
        \State \(v_2 \gets (x - x(v_0), y - y(v_0))\) for \((x, y)\in R\) closest to \(v_0\) \\\hspace*{2cm}such that \(v_2\) is not collinear to \(v_1\)
    \EndFor
\end{algorithmic}
\vspace{-8pt}
\noindent\rule{\linewidth}{0.4pt}

    \vspace{2pt} Besides the lattice structure, all other properties necessary to compile to the NA device can be retrieved from the device.
    Consequently, the resulting device model allows universal quantum ecosystems to utilize the full potential of NA devices.

    \subsection{Fidelity Improvement}\label{subsec:case study}

    To demonstrate the potential unlocked by the proposed device model, we compare two compilation scenarios:
    First, using \emph{Qiskit}~\cite{qiskit2024}---a state-of-the-art universal quantum ecosystem limited by its SC-shaped device model.
    Second, using the routing-aware NA compiler in \emph{QMAP}~\cite{stadeRoutingAwarePlacementZoned2025} as one representative NA-specific compiler that can now leverage the full device information conveyed by our model.
    For the former, limited by the SC device model, we can only specify a coupling map assuming a monolithic NA device~\cite{grahamMultiqubitEntanglementAlgorithms2022} with only local operations and without shuttling in the size of the storage zone in \cref{fig:architecture}---allowing for nearest-neighbor interactions.
    For the latter, we can fully utilize the computational capabilities of the zoned NA device~\cite{bluvsteinLogicalQuantumProcessor2023} as depicted in \cref{fig:architecture}---offering shuttling and zoned operations.
    We benchmark both settings on circuits of varying sizes taken from~\cite{quetschlichMQTBenchBenchmarking2023} as listed in the first three columns of \cref{tab:fidelity comparison}.
    For both settings, we calculate the routing overhead fidelity as the product of all operation fidelities and the coherence according to \cref{eq:coherence} on \cpageref{eq:coherence}.
    Here, we assume (local and zoned) CZ gates are executed with 99.5\% fidelity in \SI{0.36}{\micro\second}, single-qubit gates with 99.97\% fidelity in \SI{52}{\micro\second}, trap transfers (SLM \(\leftrightarrow\) AOD) with 99.9\% fidelity in \SI{15}{\micro\second}, a mean shuttling speed of \SI{0.55}{\micro\meter\per\micro\second}, and a coherence time of \(T_\mathrm{eff}=\SI{1.5}{\s}\)~\cite{linReuseAwareCompilationZoned2024,bluvsteinLogicalQuantumProcessor2023}.
    The resulting routing overhead fidelities of the SC and NA compilation are listed in the fourth and fifth columns of \cref{tab:fidelity comparison}, respectively.

    \begin{table}[t]
        \centering
        \caption{Fidelity: SC vs. NA Device Model}
        \label{tab:fidelity comparison}
        \begin{tabularx}{\linewidth}{%
            X%
            r%
            r%
            S[table-format=1.2e-1,exponent-mode=engineering,scientific-notation=true,round-mode=places,round-precision=2]%
            S[table-format=1.2e-1,exponent-mode=engineering,scientific-notation=true,round-mode=places,round-precision=2]%
            l%
        }
            \toprule
            \multicolumn{3}{l}{\textbf{Circuit}} & \multicolumn{3}{c}{\textbf{Routing Overhead Fidelity}} \\
             & {Qubits} & {Gates} & {SC Model} & {NA Model} & {Improvement} \\
            \midrule
            bv       & 32      &    46 & 0.223654                & 0.733887    & \(\times\, 3.28\) \\
                     & 64      &    94 & 0.0425464               & 0.40914     & \(\times\, 9.62\) \\
                     & 128     &   190 & 8.55762e-05             & 0.0684504   & \(\times\, 800\) \\
            ghz      & 32      &    95 & 0.27326                 & 0.469517    & \(\times\, 1.72\) \\
                     & 64      &   191 & 0.0842505               & 0.148414    & \(\times\, 1.76\) \\
                     & 128     &   383 & 0.0122803               & 0.00330402  & \(\div\, 3.72\) \\
            qft      & 8       &   142 & 0.0562852               & 0.536012    & \(\times\, 9.52\) \\
                     & 16      &   602 & 7.39139e-07             & 0.0782661   & \(\times\, 1.06\times10^{5}\) \\
                     & 32      &  2092 & {$<\!0.1\times10^{-9}$} & 0.000147453 & -- \\
            qpeexact &  8      &   136 & 0.0564316               & 0.63056     & \(\times\, 11.2\) \\
                     & 16      &   588 & 1.16278e-06             & 0.0959349   & \(\times\, 8.25\times10^{4}\) \\
                     & 32      &  2122 & {$<\!0.1\times10^{-9}$} & 8.72327e-05 & -- \\
            wstate   & 16      &    92 & 0.275245                & 0.409323    & \(\times\, 1.49\) \\
                     & 32      &   188 & 0.0762192               & 0.139574    & \(\times\, 1.83\) \\
                     & 64      &   380 & 0.0072373               & 0.00779758  & \(\times\, 1.08\) \\
            \bottomrule
        \end{tabularx}
        \vspace{-12pt}
    \end{table}

    These results clearly demonstrate that device models limited to SC features leave substantial NA potential untapped.
    When the proposed device model abstracts the full range of NA capabilities, it can bridge the gap from the hardware to specialized compilers achieving significantly higher routing overhead fidelity.
    The only exception is the \texttt{ghz} circuit with 128 qubits, where, due to the circuit's structure, long shuttling moves reduce NA performance.
    Overall, the fidelity improvements in \cref{tab:fidelity comparison} (up to \(\times\, 1.06\times10^{5}\)) highlight the significant advantages enabled by the proposed device model when employing NA devices within universal quantum ecosystems.

    \section{Conclusion}\label{sec:conclusions}

    In this work, we have identified a critical gap in universal quantum computing ecosystems: existing device models, shaped by SC technology, fail to abstract the computational capabilities of emerging technologies like NAs, including atom shuttling and zoned operations.
    This gap prevents the numerous specialized NA compilers developed in recent years from being integrated into these ecosystems, leaving substantial hardware potential untapped. %
    To close this gap, we proposed a more generic device model bridging the abstraction gap and faithfully representing NA capabilities. %
    Evaluations of an implementation of the proposed device model demonstrated fidelity improvements by a factor of up to 100,000 on a circuit with 16 qubits and 600 gates, showcasing the substantial potential unlocked by bridging the gap between NA hardware and existing software infrastructure.

    \subsection*{Acknowledgments}\label{sec:ack}

    \footnotesize
    During the preparation of this manuscript, the authors used GitHub Copilot, powered by Claude Sonnet 4.6 and Claude Opus 4.6, to improve spelling, grammar, clarity, and readability during manuscript preparation.
    Afterward, the authors reviewed and edited the content as needed.
    The authors take full responsibility for the final content.

    The project leading to this publication has received funding from the European Research Council (ERC) under the European Union’s Horizon 2020 research and innovation program (grant agreement No. 101001318), the Munich Quantum Valley, which is supported by the Bavarian state government with funds from the Hightech Agenda Bayern Plus, the Deutsche Forschungsgemeinschaft (DFG, German Research Foundation, 563436708), and the BMFTR under grant number 01MQ25001I (FullStaQD).

    \balance
    \renewcommand*{\bibfont}{\footnotesize} %
    \printbibliography
\end{document}
\typeout{get arXiv to do 4 passes: Label(s) may have changed. Rerun}